\documentclass[useAMS,usenatbib]{mn2e}
\usepackage{graphics,rotate}

\newcommand{\hMpc}{{\ifmmode{h^{-1}{\rm Mpc}}\else{$h^{-1}$Mpc}\fi}}
\newcommand{\hkpc}{{\ifmmode{h^{-1}{\rm kpc}}\else{$h^{-1}$kpc}\fi}}

\def\msun{$h^{-1}{\rm M}_{\odot}$}
\def\approxlt{\mathrel{\spose{\lower 3pt\hbox{$\sim$}}
	\raise 2.0pt\hbox{$<$}}}
\def\approxgt{\mathrel{\spose{\lower 3pt\hbox{$\sim$}}
	\raise 2.0pt\hbox{$>$}}}
\def\approxpropto{\mathrel{\spose{\lower 3pt\hbox{$\sim$}}
	\raise 2.0pt\hbox{$\propto$}}}

\title[Void Scaling]{Void Scaling and Void Profiles in CDM Models}

\author
   [Arbabi-Bidgoli \& M\"uller]
   {S.~Arbabi-Bidgoli\thanks{Email: sarbabi@aip.de}
    and V.~M\"uller\thanks{Email: vmueller@aip.de}\\
   Astrophysikalisches Institut Potsdam, An der Sternwarte 16, 
   14482 Potsdam, Germany}
\date{
      Received }

\begin{document}

\maketitle 

\begin{abstract} \noindent An analysis of voids using cosmological
$N$-body simulations of cold dark matter (CDM) models is presented.  It
employs a robust statistics of voids, that was recently applied to
discriminate between data from the Las Campanas Redshift Survey (LCRS)
and different cosmological models.  Here we extend the analysis to 3D and
show that typical void sizes $D$ in the simulated galaxy samples obey a
linear scaling relation with the mean galaxy separation $\lambda$:  $ D =
D_0 + \nu \times \lambda$.  It has the same slope $\nu$ as in 2D, but
with lower absolute void sizes.  The scaling relation is able to
discriminate between different cosmologies.  For the best standard $\Lambda$CDM
model, the slope of the scaling relation for voids in the dark matter
halos is too steep as compared to the LCRS, with too small void sizes for
well sampled data sets.

By considering a range of CDM simulations we further investigate the
scaling relation for voids within the distribution of dark matter halos
and other properties of underdense regions.  The scaling relation of
voids for dark matter halos with increasing mass thresholds is even
steeper than that for samples of galaxy-mass halos where we sparse
sample the data.  This shows the stronger clustering of more massive
halos.  Further, we find a correlation of the void size to its central
and environmental average density.  We study the evolution of the void
size distribution of dark matter halos at redshifts up to $z=3$ measuring
the sizes in comoving coordinates.  While there is little sign of an
evolution in samples of small DM halos with $v_{circ} \approx 90$ km/s,
voids in halos with circular velocity over $v_{circ}=200$ km/s are larger
at redshift $z = 3$ due to the smaller halo number density.  The flow of
dark matter from the underdense to overdense regions in an early
established network of large scale structure is also imprinted in the
evolution of the density profiles with a relative density decrease in
void centers by $\Delta (\rho /\bar{\rho}) \simeq 0.18$ per redshift unit
between $z=3$ and $z=0$.  \end{abstract}



\section{Introduction} \label{sec:introd}


Voids in the large scale structure of the universe are the focus of the
present paper.  The building blocks of the large scale structure in the
universe are superclusters, filaments of galaxies, and great walls.
These structure elements are separated by regions where the density of
visible matter is much less than the average density.  Although several
investigations of voids have been carried out (cp.  M\"uller et al.  2000
and references therein), some key questions remain open.  Voids were
found in different sizes and properties, depending on the objects
observed.  The lower space density of clusters and groups of galaxies
leads to larger voids as compared to galaxy samples.  Of particular
interest is the question of the fraction of the total volume contained in
voids.  Obviously the answer depends not only on the completeness and
quality of the data, but also on the definition used for voids.  Several
authors differ in the way that voids are identified.  While in early
papers (e.g.  Einasto et al 1983 and Oort 1983) a qualitative description
was used, later authors used spherical or spheroidal shaped regions which
were fitted into the low density regions, e.g.  Einasto et al.  (1989),
El-Ad \& Piran (2000), Plionis \& Basilakos (2001) and most recently
Hoyle \& Vogeley (2001a,b).  Aikio and M\"ah\"onen (1998) proposed a
different approach using points of maximal distance to their nearest
neighboring galaxy.  Voids detected with this algorithm are arbitrary
shaped, nearly convex regions containing no galaxies as they are formed
from a collection of subvoids.  This method yields voids which are very
similar to the results of our algorithm described below.  These methods
have the advantage of covering almost the total volume of the sample with
voids, so one can determine robustly the volume fraction that is
contained in voids of different void sizes.  Consequently the volume
fraction of voids is a function of the void size rather than a pure
number as discussed below.

Using cosmological $N$-body simulations we can explore some of the
selection effects arising while investigating data with complicated
survey geometries and object selections.  In a simulation the structure
that forms and evolves due to gravitational clustering is fully known and
accessible.  Therefore it is possible to study properties of voids such
as size, abundance and density of dark matter and small galaxy halos in
the voids and in the environment of the voids.  A drawback of this
approach is that the relation of the dark matter distribution to visible
matter such as galaxies and clusters of galaxies is still under
investigation.  A careful study of the underdense regions in dark matter
density fields and a comparison with observed voids will allow us to gain
further insight.  One challenging idea that can be realised in the
simulations is to run the clock backwards and make predictions concerning
the void properties in the past at higher redshifts.  The effect of a
possible evolution is one of our main areas of interest.

Most of our analysis is concentrated on the presently most popular set of
the cosmological parameters, the $\Lambda$CDM model.  This model includes
a cosmological constant $\Omega_{\Lambda}=0.7$ with a matter content
$\Omega_m=0.3$.  The resulting universe is spatially flat, a
preference of inflation theory.  There are presently several observations
which cover different aspects of modern cosmology and independently favor
this model among other cold dark matter scenarios.  In particular we cite
the recent measurements of the light curves of distant supernovae
\cite[]{goldh01}, measurement of the acoustic peaks in the spectrum of
the anisotropies of the cosmic microwave background \cite[]{jaffe01}, and
the evolution of the number density of galaxy clusters \cite[]{bor01}.

We propose to undertake an additional test of the popular $\Lambda$CDM
model, by comparing abundance and properties of the voids that occur in
large $N$-body simulations of this model against existing observations.
Recently in an analysis of the Las Campanas Redshift Survey (LCRS) a
reliable method of determining the voids, together with a robust
statistics in terms of median and quartile values of the size
distribution, has been developed and applied to discriminate between
different cosmological models \cite[]{mul00}.  There we studied the area
fraction of voids in an unambiguous way.  Voids in the previous
literature are often defined as empty or nearly empty regions exceeding a
certain arbitrary chosen minimum size.  To be independent of such a
minimum size, we identify voids of all sizes and let them occupy the
total volume of the survey or the sample under study.  A cumulative size
distribution shows which fraction of space is covered by voids up to a
certain size.  This cumulative size distribution has a very similar shape
in different parts of the LCRS despite differences in sampling and
possibly cosmic variance.  It represents the self similarity of the void
distribution and allows the definition of median and quartile values of
void sizes.

Previously (M\"uller et al.  2000) we showed that the median and quartile
values of the cumulative void size distribution in well defined samples
obey a linear scaling relation with the mean separation of galaxies in
the studied sample.  The slope of this relation is an important
characteristic for the clustering properties of the observed large scale
structure, and it proves to be able of discrimination between
cosmological models.  The analysis of the voids in the volume-limited
subsamples of the LCRS and the comparison to the mock samples constructed
from different CDM models independently favors a model with positive
vacuum energy $\Lambda$CDM (and/or models with a break in the primordial
power spectrum of density fluctuations, Kates et al.  1995).

Here we want to extend this analysis to 3 dimensions and make use of the
full information contained in large $N$-body simulations.  In prospect of
the upcoming surveys of the large scale structure, testing the
expectations of the models will be possible in the near future.  One of
our aims is to trace the evolution of underdense regions in the universe, 
and closely related, the evolution of the void size distribution using
simulated galaxy halos.  These were obtained in various $N$-body
simulations of the $\Lambda$CDM and the OCDM model with various box sizes
and resolutions.  Therefore we are able to explore the scope between
large simulation boxes and high spatial resolution.  We study the
connection of the void size distribution to the network of high and low
density regions of the large scale matter distribution.  Thereby we
extend the previous analytic and numeric approach of van~de~Weygaert \&
van~Kampen (1993).  Also we investigate the growth of the size of the
underdense regions as well as the evolution of their density profiles.

The present paper is organized as follows.  In the next Section we
describe the method we adopted, and in Section \ref{sec:LCRS} we present
further analysis of voids in the LCRS.  Section \ref{sec:simul} contains
a description of the CDM models and the simulations we use for our
analysis.  Section \ref{sec:3d_zero} is devoted to the study of voids and
underdensity regions at present time $z=0$ according to $N$-body
simulations.  The evolution of voids is finally examined in Section
\ref{sec:3d_evol} by investigating the cumulative void size distribution
at different epochs and the change of the corresponding averaged density
profiles of the voids.  We close with a discussion and give some
prospects for the future.


\section{Searching and quantifying voids} \label{sec:method}


The algorithm used for the identification of voids is analogous to the
one employed in our previous paper \cite[]{mul00}.  It is an adapted
version of the void search algorithm originally proposed by Kauffmann and
Fairall (1991).  The principle of the algorithm can be sketched as
follows.  A density field is realised on a high resolution grid where
each galaxy occupies one grid cell.  First the largest void is found and
the volume it occupies is assigned to it, so that no other void can enter
this region.  It proceeds to the next largest void and continues until
there remains no more empty space for voids.  Since the number of
occupied cells usually represents a negligible part of the total number
of cells, the total volume in voids occupies almost the total volume of
the survey.

Each time the void finder identifies a void, it starts with placing the
largest possible base void which is a cube of empty cells with maximal
length to the grid of the density field.  The next step of the search
goes along the six 2-dimensional faces of the base voids.  Accumulations
of empty cells are attached to the faces of the cubic shaped base void.
The condition for this is that the area covered by empty cells is larger
than two third of the face area.  Tests show that there is no significant
dependence on this condition.  This step is repeated taking each added
extension as the parent face for the next extension as long as the area
condition holds.  The procedure of adding extensions starts with placing
the largest square of empty cells on this area and attach rows of empty
cells to each side of the square.  Also the length of the row has to
exceed two third of the length of the base square.  In the next step
again the attached row is taken as the parent for the next extension
along its own edge.  After the end of this procedure, the empty cells of
the identified void are marked, so that subsequent voids will not enter
or connect to the region assigned to this void.

In case of the two dimensional surveys such as the LCRS subsamples, the
basis of the void search is a square laying in the plane of the data.
Therefore the algorithm starts with the second step, i.e.  only the part
with the squares and the rows is applied here.

Physical motivation of this procedure comes from the fact that underdense
regions in the universe tend to evolve into a nearly spherical shape with
time (cp.  e.g.  van de Weygaert \& van Kampen 1993).  While overdensity
regions decouple from the cosmic expansion and collapse to pancakes,
filaments and clusters, the underdense regions grow faster than the
background Hubble expansion.  This has the effect that the initial
geometry, which might be arbitrary, evolves to a more convex shape during
later epochs.  The above scheme takes this into account 
in two steps.  At first, the
void search starts from cubic base voids.  Then it adds extensions in a
way that approximately conserves convexity.  The algorithm identifies
neighboring voids separately if a galaxy between them stops the
extensions.  The voids are not connected by tunnel or finger like
structures.  A tunnel between two compact voids is added to the larger
void.  It may enter the smaller void region and subdivide it, but this
extension will not enclose the smaller void.  The algorithm will later
find the empty parts of the second smaller void and identify them as
separate void regions.  Therefore the present approach recognizes
large voids more robustly. 

In density fields constructed from redshift surveys or
$N$-body simulations small voids are much more abundant.  Because data
quality and sample completeness can affect each individual void, in
particular its size, shape or even existence, we need to define
statistical values, representing large, medium sized and small voids.  
In order to study the void size distribution
with samples of different density, we often dilute the simulated samples,
and the stochastic dilution mimics the effect of sparse sampling in the
observations and affects the occurrence of voids.  As it was shown by Sheth
(1996) the process of random dilution is linked to higher order
correlations of the large scale galaxy distribution, in particular in the
application to counts in cells statistics. Large voids are less
probable to be produced purely by random dilution or sparse sampling. We
regard them as more prominent for the large scale structure geometry. 


\section{2D void analysis of the LCRS} \label{sec:LCRS}


Different from our recent approach in the analysis of the LCRS, we do not
use the length of the base void as the characteristic size of the voids.
Often the volume contained in the extensions is
comparable to the base void.  Therefore we take this into consideration 
and define the diameter $D$ of a sphere
as an effective void size, that has a volume equal to the total volume of the
void, base volume plus extensions.  In case of the 2D LCRS data the
diameter of a circle is used.  Further we calculate a cumulative void size
distribution, which shows how much of the total volume of the sample is
contained in voids up to a certain size.  This cumulative distribution is
a statistically robust description of the voids as was shown in M\"uller
et al.  (2000).  As representatives for typical void sizes in the galaxy
distribution we compute the median and quartile values of the area fraction 
of voids for all samples.  In M\"uller et al.  (2000) it was
shown that these values obey a linear scaling relation $ D = D_{0} + \nu
\times \lambda$.  The mean galaxy separation $\lambda$ varies due to
different sampling fractions in the analysed data and due to different
galaxy densities for changing magnitude ranges.  We always select volume
limited data sets by imposing absolute magnitude limits.  The residuum
size $D_{0}$ -- the offset of the void size at formally zero galaxy
separation $\lambda=0$ -- describes the deviation of the scaling relation
from pure proportionality.  The galaxy separation varies due
to different galaxy densities for the absolute magnitude cuts in selecting
volume limited data sets, and in different sampling fractions in the LCRS
slices.

\begin{figure}
\resizebox{\columnwidth}{!}
{\includegraphics{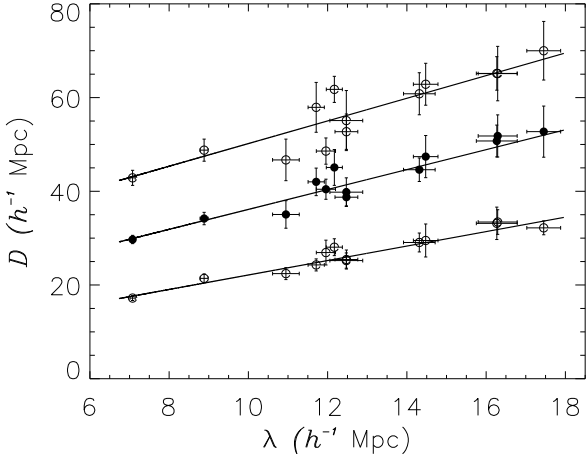}}
\caption{Median (filled circles), upper and lower quartile values (open
circles) of void sizes in the LCRS volume limited samples. The error bars 
show the variation of the results obtained by repeating the selection of 
the subsamples.}
\label{fig:lcrs1}
\end{figure}

\begin{figure}
\resizebox{\columnwidth}{!}
{\includegraphics{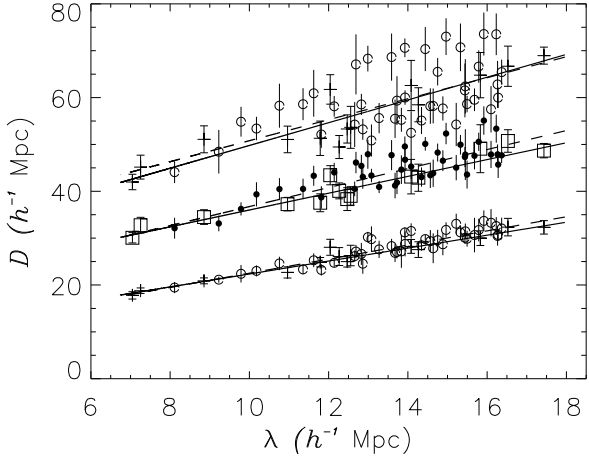}}
\caption{Median (large squares), upper and lower quartile values (diamonds) 
of void sizes in the LCRS volume limited samples with rotated grid of the
density field.  The scaling relation for these data sets are shown as
solid lines.  The quartile values for diluted subsamples of rich data 
sets ($\lambda < 12 h^{-1}$Mpc) are given as open and filled circles with 
the scaling relations as dashed lines.}
\label{fig:lcrs2}
\end{figure}

To test the robustness of the scaling relation in the LCRS found in
M\"uller et al.  (2000), here we add 3 more analyses to the original one:

\begin{enumerate} 
\item {The effects of varying sampling fraction in the data.}\\ 
\noindent 
Each of the 6 LCRS slices consists of more than 50
observational fields with varying fractions of measured galaxy redshifts.
To obtain homogeneous sampling fractions, in our previous paper we randomly
diluted all fields to the sampling fraction of the field with the minimum
sampling.  Here we repeat this approach 10 times to evaluate the
variance of the void size distribution.  The result and the fitted
scaling relations are shown in Fig.  \ref{fig:lcrs1} and  in Table
\ref{table:2d-res}.
The accuracy of
the fits and the narrowness of the error ranges are remarkable. 
The different LCRS slices
sample the voids in different parts of the universe homogeneously and
lead to stable results. This is the basis for our cosmological
conclusions.

\item {The effects of the grid orientation.}\\  
\noindent
In addition to the original paper we treated systematic errors introduced
by the grid orientation in the determination of the median and quartile
sizes $D$.  These were estimated by repeating the void search algorithm
while the volume-limited LCRS samples were rotated with respect to the
grid in steps of $10^o$.  In Fig.  \ref{fig:lcrs2} we show the scatter in
median and quartile values due to the grid orientation by the vertical
error bars on the diamonds and squares of the 14 volume limited data sets
of the LCRS.  Note that these data lie on a narrow range around the fits
to the scaling relations shown by solid lines.

\item {The effect of further random dilution in homogeneous data sets.}\\
\noindent
We also studied the scaling relation for subsamples that stem from well
sampled homogeneous datasets with mean galaxy separation $\lambda < 12
h^{-1}$Mpc, that were further randomly diluted.  Fig.  \ref{fig:lcrs2}
shows that the median and quartile void sizes of these data obey the same
scaling relation (dashed lines) as in the originally studied data sets. 
The scatter is
somewhat higher than the original data, which is due to the strong
dilution of the galaxy numbers in the restricted volumes, especially
concerning the median and the upper quartile.  We take this as an upper
bound on the variance of the void size distribution.  These data underline
that the scaling relation results from different galaxy densities in
different samples and not from different galaxy properties as magnitude
limits or galaxy types. The last point was discussed in detail in M\"uller 
et al. (2000).

\end{enumerate}

The parameters of the scaling relations for the effective void sizes $D$
are given in Table \ref{table:2d-res} first for the LCRS data using the
different random selections described in point (1) above to estimate the
error bars.  The next row in the Table contains the parameters of the
scaling relation using different grid orientations.  The third row gives the
results of a linear fit to the data including the diluted subsamples shown
in Fig.  \ref{fig:lcrs2}. The three parameter sets agree remarkably well
within the errors. However the scatter among the last sets of data is much 
larger and a $\chi^2$ test of the fits implies that only the errors contained 
in the first two lines can be regarded as reliable. For a comparison with our 
earlier paper, we also give the parameters for the scaling relation of the 
base void lengths. The absolute values are smaller because the base void 
lengths are smaller than the effective void sizes by about $\sqrt 2$. The 
new analysis better takes into account the space covered by the extensions 
to the base void, as described above in Section \ref{sec:method}.

\begin{table*}

\begin{minipage}{165mm}

\begin{center}
\caption{\label{table:2d-res} Void size scaling relation $ D = D_0 + \nu \times \lambda $ 
for the quartiles in the LCRS, $\Lambda$CDM2 halos, $\Lambda$CDM1 mock, and OCDM mock samples, 
and Poisson samples.}  

\begin{tabular}{c|ccc}

   samples              & lower quartile & median & upper quartile\\ 
\hline     
LCRS data samplings        & $ 6.7 \pm{1.1} + (1.5 \pm {0.1}) \times  \lambda    $ 
                           & $14.9 \pm{2.0} + (2.1 \pm {0.2}) \times  \lambda    $    
                           & $26.0 \pm{3.0} + (2.4 \pm {0.3}) \times  \lambda    $  \\
LCRS data for rotated grid & $ 8.5 \pm{1.0} + (1.4  \pm {0.1}) \times  \lambda    $ 
                           & $18.1 \pm{1.7} + (1.8  \pm {0.1}) \times  \lambda    $    
                           & $25.5 \pm{2.5} + (2.4  \pm {0.2}) \times  \lambda    $  \\
LCRS data for diluted samples  & $ 7.4 \pm{0.7} + (1.5  \pm {0.1}) \times  \lambda    $ 
                               & $16.5 \pm{1.3} + (2.0  \pm {0.1}) \times  \lambda    $    
                               & $28.4 \pm{1.9} + (2.2  \pm {0.1}) \times  \lambda    $  \\                          

Base void sizes (M\"uller et al.) & $ 5.7 \pm{1.6} + (0.9  \pm {0.1}) \times  \lambda    $ 
                           & $11.8 \pm{2.9} + (1.1  \pm {0.2}) \times  \lambda    $    
                           & $16.8 \pm{2.9} + (1.5  \pm {0.2}) \times  \lambda    $  \\ 

\hline
$\Lambda$CDM2 FOF-halos    & $ 7.5 \pm{1.4} + (1.5  \pm {0.1}) \times  \lambda    $ 
                           & $ 9.6 \pm{1.2} + (2.0  \pm {0.1}) \times  \lambda    $    
                           & $20.4 \pm{2.0} + (3.0  \pm {0.1}) \times  \lambda    $  \\
$\Lambda$CDM1 mocks        & $ 2.0 \pm{0.9} + (1.8  \pm {0.1}) \times  \lambda    $ 
                           & $ 6.8 \pm{2.2} + (2.9  \pm {0.2}) \times  \lambda    $    
                           & $19.7 \pm{1.9} + (2.9  \pm {0.2}) \times  \lambda    $  \\
OCDM mocks                 & $ 4.3 \pm{0.6} + (1.7  \pm {0.1}) \times  \lambda    $ 
                           & $11.2 \pm{0.9} + (2.3  \pm {0.1}) \times  \lambda    $    
                           & $16.3 \pm{1.5} + (3.2  \pm {0.1}) \times  \lambda    $  \\ 
Poisson                    & $ 1.0 \pm{0.6} + (1.9  \pm {0.1}) \times  \lambda    $ 
                           & $ 0.0 \pm{0.9} + (3.0  \pm {0.1}) \times  \lambda    $    
                           & $ 1.1 \pm{1.0} + (3.8  \pm {0.1}) \times  \lambda    $  \\                             
\end{tabular}
\end{center}
\end{minipage}
\end{table*}


\section{CDM simulations} \label{sec:simul}


To evaluate the void size distribution we performed a set of numerical
simulations of realistic CDM cosmologies.  As already shown previously
\cite[]{mul00}, the void size distribution and the scaling relation depend
on the cosmological parameters, and in particular on the power of the primordial
fluctuation spectrum on large scales as described by the shape parameter
$\Gamma = h \Omega_m$, where $h$ is the dimensionless Hubble
parameter $h = H_0/100$ km/s/Mpc and $\Omega_m$ the dimensionless
matter density.  Results that best reproduce the observed void size
distribution were obtained for a standard $\Lambda$CDM model and for a
model with a steplike primordial power spectrum, i.e.  with a similar
spectral shape as the widely discussed $\tau$CDM model.  But even there it
remained a significant discrepancy in the abundance of large voids and in
the scaling relation of the void size distribution with the mean galaxy
separation that becomes significant for the well sampled LCRS data sets.
There we could not answer the question, whether a scale dependent bias can
influence the void statistics, and in particular, what is the effect of
the suppression of the galaxy formation probability in underdense regions.
Such a suppression is sometimes denoted as large-scale bias
\cite[]{dor99}.  Furthermore, we could not answer the question for the
relation of the void statistics in the narrow LCRS slices to the void
statistics in full 3D galaxy samples.  Both questions are addressed in the
present paper using a set of realistic simulations with different
procedures for galaxy identification in high resolution dark-matter
simulations with different resolutions and box sizes.

Here we are not basically interested in discriminating between different
cosmological models, therefore we restricted us mainly to the standard
$\Lambda$CDM model with a density parameter $\Omega_m =0.3$ and a
Hubble constant $h=0.7$, and we took as a comparison model only an open
model OCDM with $\Omega_m =0.5$ and $h=0.6$.  These parameters
correspond to the models discussed in our earlier paper, and therefore,
they are well suited to answer the mentioned questions.  In Table
\ref{table:simul} we give the simulation details, in particular the
normalization by the linear mass variance on a 8 \hMpc~scale, $\sigma_8$,
the box size, and the mass and spatial resolution.

We perform two low resolution particle-mesh (PM) simulations with $256^3$
particles in large boxes of $(500\hMpc)^3$ size using $512^3$ grid cells,
they are denoted as $\Lambda$CDM1 and OCDM.  These simulations are similar
to them used previously by \cite[]{dor99} and M\"uller et al.  (2000).
With the large volume, they are well suited for discussing the relation of
voids in the effective 2D slices of the LCRS to the true 3D void
statistics.  In these models we identified galaxies with single particles
employing a nonlinear bias prescription detailed in M\"uller et al.
(2000).  In short, it employs the local density constructed from
the neighboring particles and uses a threshold bias prescription for
suppressing galaxy formation in low density regions, and a non-linear bias for
particles in high density regions to model the merging and higher dark
matter to galaxy ratio in high density regions as caused by the continuous
merging in the hierarchical clustering process, cp.  Cole et al.  (1998).

Further we employ medium resolution particle-particle particle-mesh (P3M)
simulations with the adaptive mesh refinement code of Couchmann (1991)
using also $256^3$ particles.  These simulations, denoted $\Lambda$CDM1
and $\Lambda$CDM2, use boxes of $(280\hMpc)^3$ and $(200\hMpc)^3$,
respectively, to test for the mass resolution dependence and the influence of
the upper box size on the void statistics.  The mass resolutions are 
$10^{11}$ and $4 \times 10^{10}$ \msun, respectively, and we employ a
spatial resolution of $50\hkpc$ and $40\hkpc$ in comoving coordinates, cp.
Table \ref{table:simul}.  This means we are able to resolve galaxy sized
halos with 10 - 30 particles, but we also get larger group and cluster
sized halos that enclose systems of galaxy halos being overmerged in the
medium resolution simulations.  For the question of voids, the
distribution of galaxies within groups and clusters is of minor
importance.  Therefore, we can identify field galaxy halos with systems of
about 10 particles and we study the void distributions within these
systems.

For the higher resolution simulation $\Lambda$CDM2 and $\Lambda$CDM3, we
use a standard FOF halo finder, with a dimensionless linking length (in
units of the mean particles separation) $b=0.12$ ($\Lambda$CDM2) and
$b=0.17$ ($\Lambda$CDM3).  The relatively small linking lengths
effectively collect galaxy like halos also in high density regions of the
simulation as in groups and clusters which is essential to get a realistic
indication of void boundaries in higher density regions of the cosmic web.
We impose a mass boundary of 13 DM-particles for $\Lambda$CDM2 and 8 dark
matter particles for $\Lambda$CDM3, but for many analyses we further
restrict the minimum of dark matter particles per halo.  After identifying
the halos for $\Lambda$CDM2, we check the virial theorem of the
FOF-groups, and we split off groups with high kinetic energy $T$ as
compared to the potential energy $|W|$, $T>|W|$, to separate interacting
or merging groups.  To this aim we employ a 6D-group finder with a linking
measure $\Delta r^2/r^2 + \Delta v^2/\sigma_v^2$ where $r$ and $\sigma_v$
are the effective (half-mass) radius and the velocity dispersion of the
unbound groups.  This concerns about 10\% of the halos, with many halos
having only a small number of DM particles.  The splitting off of unbound
halos adds about 2\% to the equilibrium halos due to a fixed mass boundary
for the final halos.  Thereby we get about 53000 halos resampling possible
galaxies in $\Lambda$CDM2 corresponding to a mean separation of $\lambda =
7.45\hMpc$.

For $\Lambda$CDM3, we select halos at redshifts $z=0, 1, 2, 3$ to check for
the evolution of the void sizes.  We take all halos selected with a linking length 
$b=0.17$,
finding about 160 000 objects with a mass limit of 8 dark matter particles,
i.e.  with a mean interparticle separation of $\lambda=3.7 \hMpc$.  At
$z=3$ this grows to $\lambda=4.4 \hMpc$ due to the smaller number of halos
at this redshift.  We also employ a minimal circular velocity $v_{circ} >
200$ km/s for identifying normal mass galaxy halos which leads to about
23000 objects with a mean separation of $\lambda = 7\hMpc$ at $z=0$.  The
mean separation between halos grows at higher redshift, corresponding to
$\lambda=7.8 \hMpc$ at $z=1$, $\lambda=9.4 \hMpc$ at $z=2$, and $\lambda=13
\hMpc$ at $z=3$ due to the mass increase in single halos by merger
precesses and continuous mass accretion.

As shown in M\"uller et al.  (2000) for the nonlinear biased mock
catalogues, the halos reproduce well the two-point correlation function of
galaxies in the LCRS in the redshift space \cite[]{tuc97}.  Therefore they
are reasonable test objects tor the void statistics in the
underlying CDM models.

\begin{figure}
\resizebox{\columnwidth}{!}
{\includegraphics{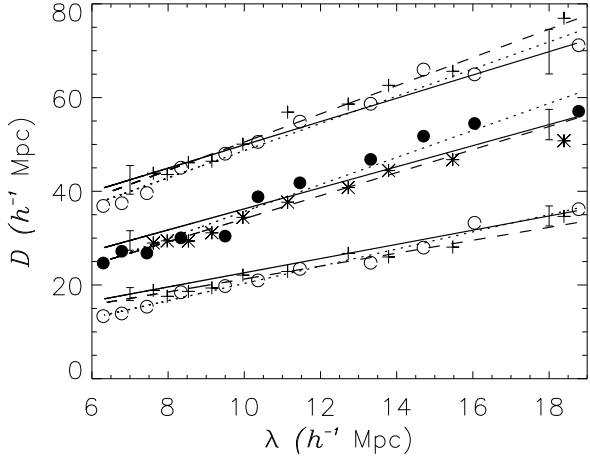}}
\caption{Scaling relation for median and quartile values of the void sizes 
of mock galaxy samples of the OCDM simulation in a 500$h^{-1}$Mpc box 
(filled and open circles and dotted lines) and halos in the $\Lambda$CDM2 
simulation using a 280$h^{-1}$Mpc box (stars, crosses, and dashed lines), 
both in a geometry similar to the volume-limited sub-samples of LCRS 
(solid lines with error bars).}
\label{fig:lvsdata}
\end{figure}

The key questions of the simulations is whether they can reproduce the void
size distribution, and in particular, the slope of the scaling relation for
randomly diluted model galaxy samples. This characteristic measures the
clustering properties in different parts of the universe observed by the
LCRS.  As it was shown in the previous paper, this slope is smaller for the
LCRS samples than for the mock samples of N-body simulations, i.e.  the
median and upper quartile void sizes determined in the dense parts of the
LCRS are larger.  This was shown using mock samples having the same
geometry as the LCRS samples.  Especially in the well sampled parts the
mock samples are not able to reproduce large voids.  We complement this
analysis by studying friend-of-friend dark matter halos of the
$\Lambda$CDM2 simulation (cp.  Table \ref{table:simul}).  In Fig.
\ref{fig:lvsdata} we show that the scaling relation of voids in the OCDM
mock samples (dotted lines) has almost the same steep slope as the scaling
relation of halos in the $\Lambda$CDM2 simulation (dashed line, cp.  also
Table \ref{table:2d-res}).  In particular, it is significantly steeper than
that from the LCRS (solid lines).  The analysis of the halo samples was
performed by constructing mock samples in a slice geometry 
as in the LCRS data, and
we did a 2D void search as described in the last section.  
To resume, we cannot reproduce the typical voids in well sampled
parts of the LCRS as shown on the left of Fig.  \ref{fig:lvsdata}.

\begin{figure}
\resizebox{\columnwidth}{!}
{\includegraphics{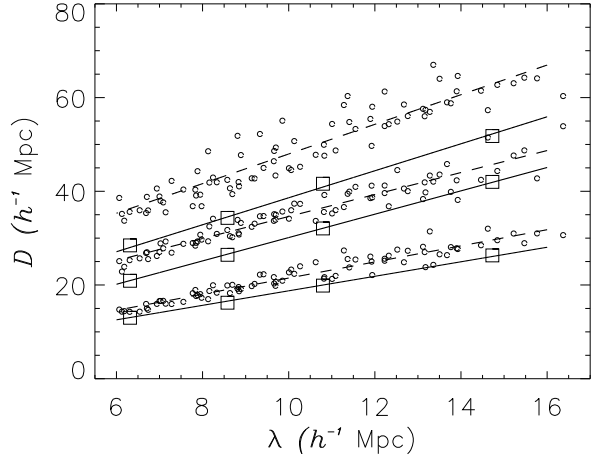}}
\caption{Comparision of the scaling relation in 2D and 3D for mock galaxy
samples of the OCDM simulation in a 500$h^{-1}$Mpc box. The 2D results are 
based on a geometry similar to LCRS volume-limited samples (small circles 
and dashed lines). The 3D analysis leads to smaller values of the medium 
and quartile void sizes (large squares and solid lines), but the slope of 
the scaling relation is similar.}
\label{fig:mock2d3d}
\end{figure}

\begin{table*}

\begin{minipage}{100mm}

\begin{center}
\caption{\label{table:simul} Cosmological $N$-body simulations}  

\begin{tabular}{c|cccccc}
  model   & $\Omega_m$ &  h & $\sigma_8$ & box size & mass resolution & spatial resolution \\ 
  \hline    
  $\Lambda$CDM1 & 0.3  & 0.7 & 0.87 & $500\hMpc$ & $6 \times 10^{11}$ \msun & 1.0  \hMpc \\
  $\Lambda$CDM2 & 0.3  & 0.7 & 0.91 & $280\hMpc$ & $10^{11}$ \msun          & 0.05 \hMpc \\
  $\Lambda$CDM3 & 0.3  & 0.7 & 0.87 & $200\hMpc$ & $4 \times 10^{10}$ \msun & 0.04 \hMpc \\
  OCDM          & 0.5  & 0.6 & 0.80 & $500\hMpc$ & $10^{12}$ \msun          & 1.0  \hMpc \\
\end{tabular}
\end{center}

\end{minipage}
\end{table*}

The values found for the slopes of the scaling relations for the quartile
and median values are given in Table \ref{table:2d-res}.  It shows that
the OCDM and $\Lambda$CDM1 mock samples and the $\Lambda$CDM2 FOF halos,
that were constructed in different simulations, have similar scaling
relations, in particular similar slopes.  In all samples, we notice a
statistically significant discrepancy to the LCRS.


\section{Voids in 3D at redshift zero} \label{sec:3d_zero}


While in Section \ref{sec:method} mock and halo samples with a slice geometry
similar to the LCRS were analyzed, now we proceed to 3-dimensional samples
obtained from simulations.  As a typical example results from the OCDM
simulation in a 500$h^{-1}$Mpc box are shown in Fig.  \ref{fig:mock2d3d}.
In 10 different cuts through the box, diluted mock samples with LCRS
geometry were constructed.  Their scaling with the mean separation is
indicated by the dashed lines.  Additionally the 3-dimensional algorithm is
applied to the complete box.  The full sample was also diluted in this case
to investigate the behavior with lower sample density.  Not only do the
voids obey the scaling relation, but also the scatter is typically much
lower.  The scaling relations show the same slope, while there is an
overall offset in the size.  We regard this difference as a feature of the
two dimensional slices.  By observing only thin sheets in the underlying
3-dimensional galaxy sample, as it is the case with the LCRS slices, the
effective size of voids {\em increases} due to missing galaxies in the
bounded region of the slices.  The increase in the absolute value of the
scaling relation obviously depends on the mean depth of the slice in
comparison to the mean void parameters, and we cannot connect it with any
fundamental property of the galaxy distribution.  The observed void sizes
become on average 20 - 30\% smaller in 3D, but the slope of the scaling
relation of void sizes with the mean galaxy separation remains constant and
represents an important characteristics of the hierarchy of the galaxy
distribution in the cosmic web.

An important question to be addressed while studying void sizes is the
dependence on the mass of the objects which represent the structure under
analysis.  From elementary considerations typical voids expected within
the distribution of dwarf galaxies as an example must be different in size
from the voids present in the large scale structure built by galaxy
clusters, as an extreme opposite example.  This is firstly due to the
different number densities of these objects, and secondly because smaller
objects are distributed more smoothly on average than large objects.  Both
of these properties affect the sizes of the voids, and they can be
quantified by the scaling relation.

\begin{figure}
\resizebox{\columnwidth}{!}
{\includegraphics{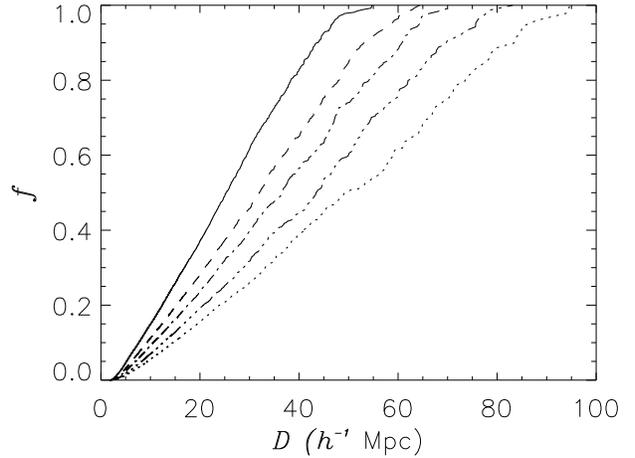}}
\caption{The cumulative void size distributions in the  $\Lambda$CDM2 
simulation in a 280$h^{-1}$Mpc box. We show the size distributions for 
dark matter halos with masses increasing from left to right, from 
$1.3 \times 10^{12}$ \msun (solid line), $3.2 \times 10^{12}$ \msun 
(dashed line), $6.0 \times 10^{12}$ \msun (dashed dotted line), 
$9.0 \times 10^{12}$ \msun (dashed three-dotted line) to 
$1.5 \times 10^{13}$ \msun (dotted line).}
\label{fig:halo_g}
\end{figure}

In M\"uller et al.  (2000) we studied a possible dependence of median void
sizes on the absolute magnitude.  The magnitude range of galaxies in the
LCRS varied between -20 and -21.  This is a small range and as such we
could not find any significant change in the sizes of voids.  Here we
study the dependence of void sizes on the FOF halo mass.  Fig.
\ref{fig:halo_g} shows that the cumulative void size distribution bends
towards larger void sizes with growing mass of the halos.  A $\Lambda$CDM
simulation in a 280$h^{-1}$Mpc box with a mass resolution of
$10^{11}M_\odot$ was used.  The lower cutoff in the mass range was set to 
$1.3 \times 10^{12}M_\odot$ for the most complete sample and to $ 1.5 \times
10^{13}M_\odot$ for the one with the lowest space density of halos.  In
the first case the sample contains all objects with a mass comparable to a
milky way galaxy and higher.  In the latter case the objects that remain
in the sample are comparable to small galaxy groups.  Because the number
density of the identified halos decreases with growing mass, 
the structure of the samples is 
dominated by halos with masses near the lower end.  Within the mass range
of halos under investigation the median value of the void sizes increases
from 25$h^{-1}$Mpc to twice this value.  Obviously the void size
distribution looks remarkably similar, cp.  also M\"uller et al.
(2000).

\begin{figure}
\resizebox{\columnwidth}{!}
{\includegraphics{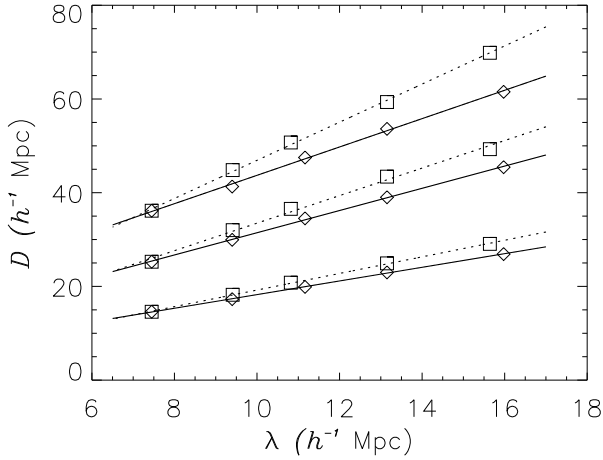}}
\caption{The median, upper and lower quartile
values of the void size distributions of DM halos with growing mass
as described in Fig. \ref{fig:halo_g} (squares and dotted lines), 
compared to diluted samples using the lowest mass threshold 
(diamonds and solid lines).}
\label{fig:gvsdil}
\end{figure}

As a comparison of the scaling relation from sparse sampling and
increasing mass, we construct randomly diluted samples with average mean
separations $\lambda$ similar to the halo samples with different masses.
The diluted sets contain a mixture of high and low mass halos.  Because of
the higher abundance of the low mass halos, they dominate the diluted
samples.  Althouth dilution increases the void sizes, the voids in the
diluted mixed samples are significantly smaller.  As it is shown in Fig.
\ref{fig:gvsdil} the difference grows to 20\% for the sample with the 
largest
halo masses.  This effect is expected and easily explained by the picture
of gravitational instability.  Larger mass halos 
are more clustered according to 
the higher amplitude of the 2-point correlation function of galaxy groups
as compared to galaxies.  Large and massive halos reside in the nodes of
the large scale structure network, while the less massive ones are still
streaming along pancakes and filaments towards these nodes.  Their
distribution is more homogeneous and less clustered.  Table
\ref{table:3d-res} compares the parameters of the scaling 
relation for these two
differently selected series of halo samples. 
A similar change of the void size was found in Little
\& Weinberg (1994), but they did not derive the scaling relation of the
void sizes under random dilution studied here.  There were found larger
voids if a biasing prescription for galaxy formation was applied.  With
the present approach we quantify this relation of object masses and void
sizes.

\begin{table*}

\begin{minipage}{165mm}

\begin{center}
\caption{\label{table:3d-res} The scaling relation in 3D
for the lower quartiles, medians and upper quartiles of void size
distributions in randomly diluted OCDM and FOF-halo samples and in halo
samples with increasing mass.}

\begin{tabular}{c|ccc}

   samples                           & lower quartile & median & upper quartile\\ 
  \hline    
OCDM mocks, randomly diluted         & $ 3.3 \pm{0.2} + (1.5  \pm {0.02}) \times  \lambda    $ 
                                     & $ 5.2 \pm{0.1} + (2.5  \pm {0.01}) \times  \lambda    $    
                                     & $ 9.8 \pm{0.5} + (2.9  \pm {0.06}) \times  \lambda    $  \\
$\Lambda$CDM halos, randomly diluted & $ 3.7 \pm{0.2} + (1.5  \pm {0.02}) \times  \lambda    $ 
                                     & $ 7.7 \pm{0.4} + (2.4  \pm {0.04}) \times  \lambda    $    
                                     & $13.5 \pm{0.8} + (3.0  \pm {0.07}) \times  \lambda    $  \\      
$\Lambda$CDM halos, increasing mass  & $ 1.5 \pm{0.3} + (1.8  \pm {0.02}) \times  \lambda    $ 
                                     & $ 4.1 \pm{1.5} + (2.9  \pm {0.13}) \times  \lambda    $    
                                     & $ 6.2 \pm{0.8} + (4.1  \pm {0.07}) \times  \lambda    $  \\                        
\end{tabular}
\end{center}
\end{minipage}
\end{table*}

\begin{figure}
\resizebox{\columnwidth}{!}
{\includegraphics{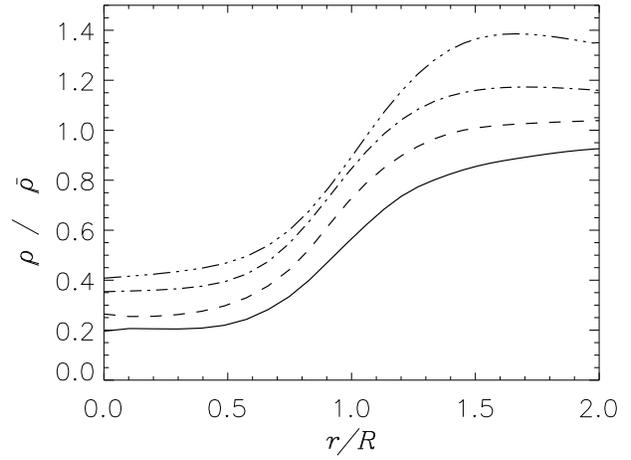}}
\caption{Averaged density profiles of under dense regions in a 
$\Lambda$CDM2 simulation with 200$h^{-1}$Mpc box size. The average 
is taken in size ranges 36$h^{-1}$Mpc - 50$h^{-1}$Mpc (solid line),
24$h^{-1}$Mpc - 36$h^{-1}$Mpc (dashed line), 16$h^{-1}$Mpc - 
24$h^{-1}$Mpc (dashed-dotted line) and 10$h^{-1}$Mpc - 16$h^{-1}$Mpc 
(dashed-3dotted line).}
\label{fig:prof-z0}
\end{figure}

An additional study illustrates the relation of the void sizes to their
environmental and central density.  For this purpose a smoothed density
field was constructed in the dark matter distribution of the $\Lambda$CDM3
simulation in a 200$h^{-1}$Mpc box.  In the first step the dark matter
particles were distributed onto a grid with 0.5 $h^{-1}$Mpc per cell,
using the cloud-in-cell method.  After smoothing the density field with a
Gaussian function ($\sigma = 1$ $h^{-1}$Mpc) a threshold overdensity of
$\delta_{th}=5.0$ was applied.  Every grid cell with a density below this
value was regarded as containing no significant structure.  After finding
the voids their density profiles were computed by averaging in concentric
spheres around the center of the base voids.  The density
profiles as shown in Fig.  \ref{fig:prof-z0} were obtained by averaging in
size ranges between $(36 - 50) h^{-1}$Mpc, $(24 - 36) h^{-1}$Mpc, $(16 -
24) h^{-1}$Mpc and $(10 - 16) h^{-1}$Mpc.  To obtain the average between
voids of different sizes we use a relative radial coordinate $r/R$, where
$r$ is the distance from the void center and $R$ is the corresponding 
effective void
radius defined above.  For the study of the environment of the voids we
extend the averaging to relative radial coordinates larger than unity.
They are calculated up to a sphere with $r=2R$, the last sphere contains a
volume eight times larger than the void.

Two effects are obvious in the profiles shown in Fig.  \ref{fig:prof-z0}.
The average underdensity at the center of voids depends on the void size.
On average the central density of small voids is higher than that of large
voids.  This correlation reflects the inversion of gravitational
clustering.  While high peaks in the density fluctuations evolve to the
knots in the large scale structure, large voids are formed in regions with
a primordial large scale underdensity.  Also the profiles show the
tendency that the environment of large voids is underdense, while small
voids will form both in underdense and overdense regions.  This is 
typical for the hierarchical clustering scenario.

\section{Evolution of Voids } \label{sec:3d_evol} 


A study of the void evolution using an adhesion model of structure
formation was carried by Sahni et al.  (1994).  There a correlation of the
void size with the value of the primordial gravitational potential was
suggested, and it was derived a redshift dependence of void sizes,
$D(z)=D_{0}/(1+z)^{1/2}$.  More recently, Friedmann and Piran (2000)
proposed a model that combines the growth rate of negative density
perturbations with a lower probability of galaxy formation in these
underdense regions.  According to this model a typical void with a size of
20$h^{-1}$Mpc corresponds to a 3$\sigma$ perturbation so that they occur
very rarely.  This simple model can explain relatively large voids in the
case of a $\Lambda$CDM universe, but it still fails to reproduce the
occurrence of as large voids as observed in the LCRS \cite[]{mul00}, or in
the ORS and IRAS Galaxy catalogues, \cite[]{elad00}.

Here we make use of the $\Lambda$CDM3 simulation in a 200$h^{-1}$Mpc box.
In addition to the resulting dark matter distribution at present epoch, we
stored the snapshots of the evolution at $z=1$, $z=2$ and $z=3$.  With the
FOF-algorithm described above, gravitationally bound dark matter halos of
were identified at each time step to study the effects of void size
evolution.  A minimum of 8 dark matter particles was chosen as a
requirement for the halo identification, but we keep in mind that 
these low mass halos are not well resolved and quite
unstable to disruption in the tidal fields of neighboring halos.  These 
halos have a minimum mass of $ 3 \times 10^{11}M_\odot$.
Furthermore we constructed halo samples with a minimum circular velocity
v$_{circ}= \sqrt{G M /r_{vir}} =200$ km/s within the virial radius
$r_{vir}$. 

\begin{figure}
\resizebox{\columnwidth}{!}
{\includegraphics{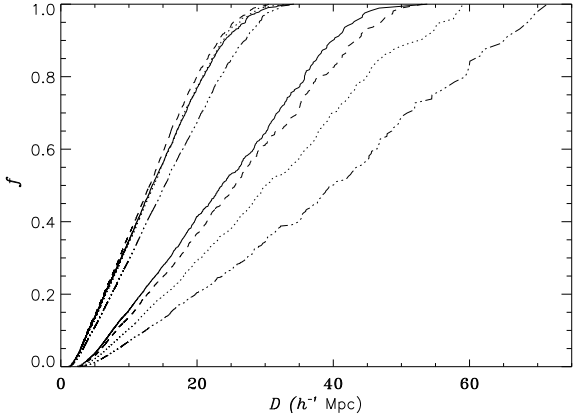}}
\caption{Cumulative size distribution of voids among
dark matter halos in a $\Lambda$CDM simulation with 200$h^{-1}$Mpc size.
The solid lines refer to z=0, dashed, dotted and dashed-dotted lines
refer to z=1, z=2, and z=3, respectively. The distributions to the left 
are obtained in the case of complete halo distributions, and the others
belong to the case with a lower cutoff in the circular velocity 
$v_{circ}=200$ km/s.} 
\label{fig:cum-evol}
\end{figure}

As Fig.  \ref{fig:cum-evol} shows there is little sign of evolution up to
a redshift of $z=2$ if we consider the complete samples.  This means that
void sizes grow in the same rate as the Hubble expansion.  The network of
high and low density regions in the universe has been formed in an early
epoch of the cosmic evolution.  But on the other hand a significant
evolution of the void distribution is observed when we consider halos
larger than a certain mass or circular velocity.  This implies that after
the early setup of the large scale network the matter flows along the
walls and filaments, continues to merge and forms larger objects with
time.  At the redshift $z=1$, the evolution of the number density is
already visible in the void distribution.  At redshifts higher than unity,
large halos are still in the stage of forming.  Their number density is
significantly lower, and the voids in the spatial halo distribution are
increasing in comoving coordinates.  The formation of more massive 
objects leads to a
subdivision of voids as we approach redshift $z=0$.

\begin{figure*}
\resizebox{18cm}{!}
{\includegraphics{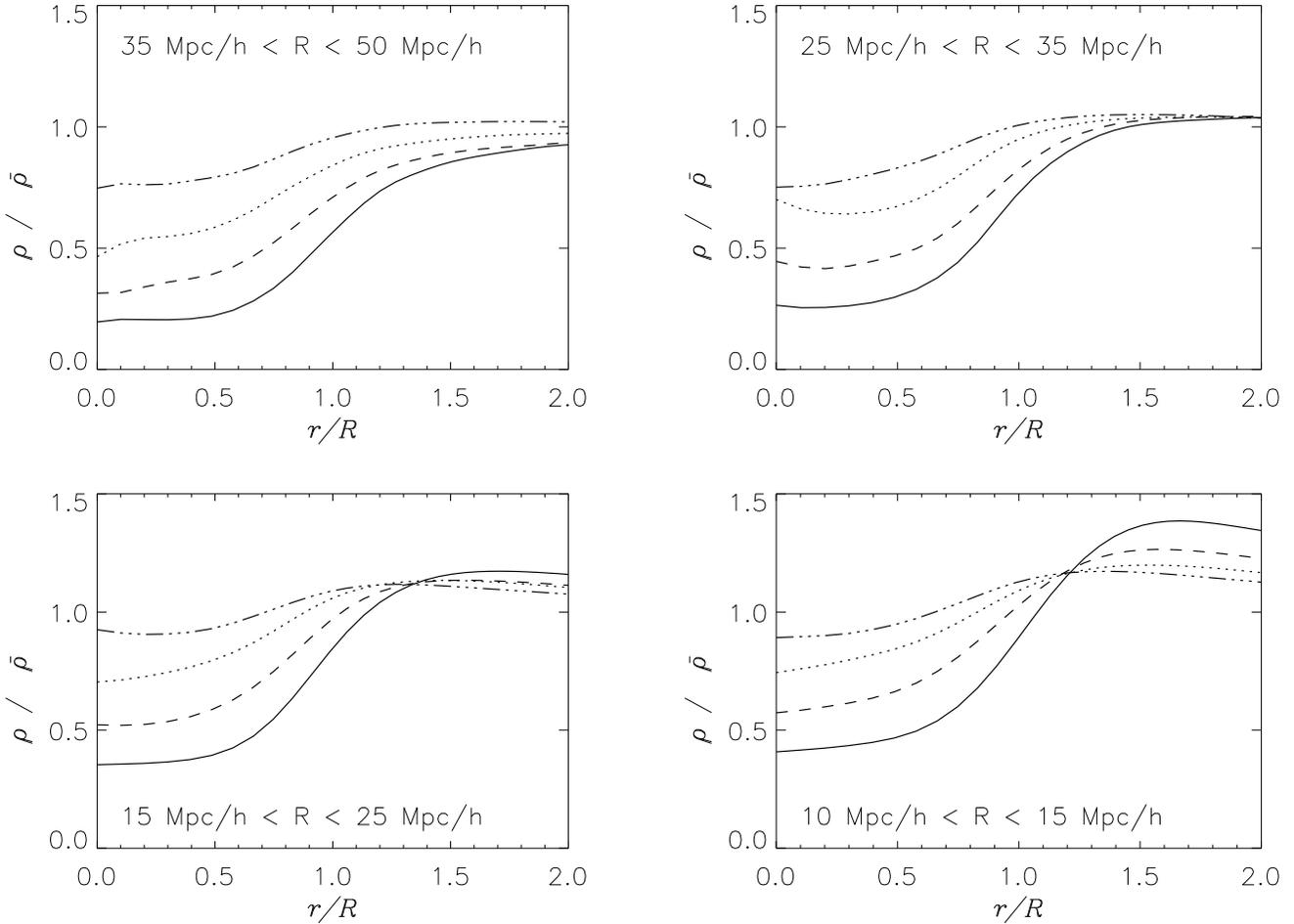}}
\caption{The evolution of average density profiles for voids in the 
same ranges as in Figure \ref{fig:prof-z0}. Each panel shows the 
evolution in one size range, from top left the largest voids to lower 
right the smallest ones. The profile curves in each panel correspond 
to different epochs, z=0 (solid lines), z=1 (dashed lines), z=2 (dotted 
lines) and z=3 (dashed-3dotted lines) and illustrate the evolution in 
each size range.} 
\label{fig:p-evol}
\end{figure*}

The evolution of the underdense regions is accompanied by a flow of dark
matter from low dense to high dense regions.  Here we study the evolution
of the density profiles as described above.  The resulting averaged density
profiles are shown in Fig.  \ref{fig:p-evol} at different redshifts.  The
density profiles develop distinctly for different sizes.  The value of the
average density in the central parts decreases linearly in all panels with
$\delta=0.18$ per redshift unit.  The upper left panel of Fig.
\ref{fig:p-evol} corresponds to the largest underdense regions.  Different
from the other size ranges, the average density in the environment of such
voids decreases during the evolution from $z=3$ to $z=0$.

In the upper right and lower left panel of Fig.  \ref{fig:p-evol}, which
show the profiles of median size underdense regions, the average density
of the void environment remains nearly constant and roughly at mean cosmic
density.  In the lower right panel, which corresponds to smaller voids,
the average density in the environment grows with redshift.  Fig.
\ref{fig:p-evol} illustrates the flow of dark matter from low into high
density regions mainly surrounding the small voids.  A correlation between
regions with positive initial gravitational potential and underdense
regions was also found in \cite[]{mad98}, in good agreement with our
interpretation. There it was also found a decreasing average density in 
the underdense regions.

\section{Discussion} \label{sec:disc} 


The present paper includes an extension of our previous analysis of the
void size distribution in the 2D samples of the LCRS.  By comparison of
large 3D boxes with mock 2D slices in similar geometry as the LCRS we can
extend our previous results.  Although there is more scatter and
uncertainty in the statistical measures of the void sizes in 2D, there is
evidence that these results are physically relevant for the complete
spatial distribution of galaxies.  From the simulations we find a strong
hint that the observed scaling relation of effective 
void sizes with the mean galaxy
separation in the LCRS samples has a counterpart in the 3D structure. 
The effective size was defined both in 2D and 3D as the diameter of the 
equal sized circle or sphere as the voids. The
two differ only by a nearly constant offset, i.e.  the 3D void sizes 
that occupy the same volume fraction (or area for 2D) are about 20\% 
smaller. We
test the robustness of the scaling relations of the LCRS by 
evaluating the variances that
the differently selected homogeneous samples, the differently oriented 
grids, and the dilutions introduce into the results.  Other possible sources of
systematic errors, such as the influence of the selection function and
geometry and boundary effects, were investigated by our previous paper.
The uncertainty due to these effects was found to be less than 10\%.

The next part of our analysis is devoted to the question of how the void
sizes in the spatial distribution of some objects depend on the type of
objects.  Here we study the dependence for halos of milky way mass and
above.  Below this mass range there might be some inconsistency of the
galaxy population in voids between the observational data and models of
galaxy formation \cite[]{peeb01}.  As we restrict our samples to higher
mass halos, it is naturally to expect that the identified voids are
larger, just because the number density of halos decreases and the mean
halo separation becomes larger.  An important result obtained with our
analysis is that larger voids sizes are found for halos with larger mass.
This effect is shown by comparing the scaling relation for samples with
increasing halo masses and diluted subsamples of the original sample with
equal mean density.  In the latter case the voids are smaller and the slope
of the scaling relation is shallower because a mixture of halos with
different masses is less clustered than the comparable sample of only high
mass halos.  Also in this respect the scaling relation of void sizes
proves to be a good measure for the fundamental clustering properties of a
sample.  Our new statistics extends the void probability function that was
shown to provide a generating function for the set of higher order
correlation functions (White 1979, Sheth 1996). It is more sensitive to 
the large scales in the cosmic network of galaxy structures. 

Furthermore we discussed the void density profiles.  These were estimated
by averaging in spherical shells around the void centers.  They show
characteristics which are typically different for large and small voids.
The central and environmental density of large voids are significantly
lower than small voids.  The underdensity around a large void encompasses
a larger volume than the void itself, while the environment of small voids
represents in the average an overdense ridge.  We also examined the
evolution of void profiles up to a redshift $z=3$.  The central
underdensity decreases linearly with $\delta=0.18$ per redshift unit.  Also
the environmental underdensity of large voids becomes deeper during the
evolution.  For average sized voids the environment density remains
constant near the mean matter density, while the overdensity ridge of
small voids grows.  Regarding the size distribution and volume fraction of
voids we find little sign of evolution in catalogues containing small dark
matter halos.  However if we restrict the samples to larger mass halos
comparable to the milky way, there is a significant evolution in their
number density and consequently in the void distribution.  As they
continue forming they subdivide existing voids and shift the void size
distribution towards lower sizes in comoving coordinates.

The exploration of the void size evolution and of the void scaling
relation in new redshift catalogues and in more realistic model galaxy
distributions than considered here represents a challenging prospect for
the future.  It should allow a quantitative understanding of the large
scale environmental dependence of galaxy formation and its bias that
should be basic for resolving the discrepancy of the scaling
relation of the void size distribution function in the LCRS and in CDM
models.  In this way we can find an answer to the question, whether this
discrepancy is a fundamental problem for the standard $\Lambda$CDM model
or whether it can be explained by a more realistic description of the
galaxy formation than it was possible by a nonlinear bias and by our
friend-of-friends halo finder.


\section*{Acknowledgements}

Sincere gratitude is due to the anonymous referee for useful and detailed
comments. S.A.B. acknowledges support from the German Science foundation 
(DFG - Mu 6/1). We employed the AP3M code for the simulations kindly made 
public by H. Couchman.


\bibliographystyle{mnras}
\bibliography{vscale}

\end{document}